\documentclass{article}
\usepackage[]{authblk}
\usepackage[colorlinks=true, citecolor=blue, urlcolor=blue]{hyperref}
\usepackage[english]{babel}
\usepackage{url}
\usepackage[numbers,sort&compress]{natbib}
\newcommand{\mockalph}[1]{}
\usepackage{amsmath,amssymb,amsthm}

\emergencystretch=\maxdimen
\begin{document}

\title{SurpriseMe: an integrated tool for network community structure characterization using Surprise maximization}
\author{Rodrigo Aldecoa and Ignacio Mar\'in\,\footnote{to whom correspondence should be addressed}}
\affil{\small Instituto de Biomedicina de Valencia, CSIC}
\date{}

\maketitle

\section*{Abstract}
Detecting communities, densely connected groups may contribute to unravel the underlying relationships among the units present in diverse biological networks (e.g., interactome, coexpression networks, ecological networks, etc.). We recently showed that communities can be very precisely characterized by maximizing Surprise, a global network parameter. Here we present SurpriseMe, a tool that integrates the outputs of seven of the best algorithms available to estimate the maximum Surprise value. SurpriseMe also generates distance matrices that allow to visualize the relationships among the solutions generated by the algorithms. We show that the communities present in small and medium-sized networks, with up to 10.000 nodes, can be easily characterized: on standard PC computers, these analyses take less than an hour. Also, four of the algorithms may quite rapidly analyze networks with up to 100.000 nodes, given enough memory resources. Because of its performance and simplicity, SurpriseMe is a reference tool for community structure characterization.

\subsubsection*{Availability and implementation}
The source code is freely available under the GPL 3.0 license at \url{http://github.com/raldecoa/SurpriseMe/releases}. SurpriseMe compiles and run on any UNIX-based operating system, including Linux and Mac OS/X, using standard libraries.

\subsubsection*{Contact} \href{imarin@ibv.csic.es}{imarin@ibv.csic.es}

\section{Introduction}
Complex networks are extensively used for representing interactions among elements of a system. This approach is particularly useful in biology: analyzing networks provides relevant information in fields such as genetics \citep{Costanzo_et_al}, neuroscience \citep{Bullmore_and_Sporns}, ecology \citep{Bascompte_et_al}, systems biology \citep{Barabasi_and_Oltvai} or proteomics \citep{Schwikowski_et_al}. An interesting property of these networks is the fact that related nodes tend to create tightly knit groups, usually known as \textit{communities}. By unraveling the close relationships among certain units, community structure characterization improves our understanding of the system as a whole.

In the last years, many strategies have been devised to detect the optimal division into communities of a network. However, none of them alone is able to achieve high quality solutions in all kind of networks \citep{Schaub_et_al, AM13a, AM13b}. In recent works, we demonstrated that Surprise (S) \citep{Arnau_et_al, AM10, AM11} is an effective measure to evaluate the quality of any partition of a network \citep{AM11, AM13a, AM13b}. In several complex benchmarks, composed of networks with very different structures, it has been shown that the partition of maximum S corresponds to the real community structure, with a minimal/null degree of error \citep{AM11, AM13a, AM13b}. Although a simple algorithm to maximize S has not been yet devised, it was shown that combining the output of seven high-quality algorithms, always choosing the one that provided the maximum value of S, was sufficient to solve the structure of the networks tested. These algorithms were CPM \citep{CPM}, Infomap \citep{Infomap}, RB \citep{RB}, RN \citep{RN}, RNSC \citep{King_et_al}, SCluster \citep{AM10} and UVCluster \citep{Arnau_et_al, AM10}. 

In this article we present SurpriseMe, a tool integrating those seven algorithms. SurpriseMe accelerates the research process by simply accepting a network as input, running internally all those algorithms and outputting their solutions together with their Surprise values. SurpriseMe also calculates distances among the solutions provided by the algorithms, an information that allows to understand how congruent they are \citep{AM13b}.

\section*{Methods}
\subsection*{SurpriseMe: S maximization and distances among solutions}
SurpriseMe requires as an input a text file indicating the list of links that characterize the network. Each line of the file contains a link, represented as a pair of nodes separated by a tab or space character. From this text file, the software provides the different programs with the appropriate input files.

As indicated above, SurpriseMe analyses are focused on maximizing the Surprise (S) parameter. Given a partition of a network into communities, S calculates the unlikeliness of finding the observed number of intra-community links in a random network. It is based on a cumulative hypergeometric distribution \citep{Arnau_et_al, AM11}:	

\begin{equation}
  S = -\log \sum_{j = p}^{\min(M,n)} 
    \frac{\displaystyle {\binom{M}{j}} 
          {\binom{F - M}{n - j}}}
         {\displaystyle {\binom{F}{n}}}
\end{equation}

where F is the maximum possible number of links of the network, n is the actual number of links, M is the maximum possible number of intra-community links and p is the actual number of links within communities. SurpriseMe calculates either the S values for the seven algorithms or of a subset of  them chosen by the user, establishing which one is the best, maximum one. The program also compares all the solutions using either the Variation of Information (VI) \citep{Meila} or the value that corresponds to (1 - NMI), where NMI means Normalized Mutual Information \citep{Danon_et_al}. In both cases, distance = 0 means that two solutions are identical, and the greater the value, the more different are two partitions. Details of the differences of using VI versus NMI can be found in \citep{AM12, AM13a, AM13b}. The program also estimates the distances to two artificial solutions called ``One'' (all units of the network are in one community) and ``Singles'' (each node belongs to a different community). The distances to these two solutions provide additional clues about how each algorithm is behaving \citep{AM13b}, All these distances are saved into two distance matrix files (one for VI, another one for 1-NMI) that can be directly imported into MEGA \citep{MEGA}, a popular free software which allows an easy visualization of the hierarchical relationships among the different solutions, as shown in \citep{AM13b}.

\subsection*{Performance}
Given the substantial complexity of the algorithms involved, the current version of SurpriseMe is most useful for networks of small to medium size, typically up to 10.000 nodes. We established the performance of the software by analyzing two types of standard benchmarks. One consisted of networks based on a Relaxed Caveman (RC) configuration \citep{Watts} with 10\% rewiring, which means that well-defined communities are present \citep{AM11, AM13a, AM13b}. The second was a set of Erdös-Rényi (ER) random graphs \citep{Erdos_Renyi}, essentially without community structure. This last benchmark provides an estimate of the maximum time and resources required.

Both with RC and ER structures, networks with up to 10.000 nodes are analyzed by the 7 algorithms in less than an hour using a conventional desktop PC, consuming less than 1 GB of memory. However, larger networks require more powerful hardware and it may be then advisable to switch off the most time- and resource-consuming programs, which are RN, SCluster and UVCluster. Although this obviously may limit S maximization in some cases, close to optimal solutions are generally provided by the four remnant programs in ways that are moreover complementary (i.e., they work optimally in different network structures; see \citep{AM13a, AM13b}), so their combination will still generate either very high or maximum S values. With all the programs, we have estimated that a RC network of 50.000 nodes requires 140 hours of analysis and around 60 GB of memory. This is reduced to 40 minutes and 14 GB of memory (RC structure) or 8 hours and 39 GB of memory (ER configuration) if only the four fastest programs are used. For a RC network of 100000 nodes, we have determined that the four fastest algorithms take 3 hours and 30 GB of memory in RC benchmarks, which goes up to 21 hours and 66 GB of memory for ER networks.

\section*{Summary}
Only few researchers have the time and skills to select, download, compile and run multiple community detection algorithms. SurpriseMe allows to very simply run a set of state-of-the-art algorithms and determine which one generates the best Surprise value, i.e., the best partition of the network. It also provides the user with distance matrices (with VI, 1-NMI values) that may help to understand how the solutions of the different algorithms compare. Very simple to use, it only needs as input a file containing the network to analyze. The well-established power of this type of analysis together with the simplicity of its use, make SurpriseMe an excellent tool for characterizing the community structure of complex networks.

\subsection*{Acknowledgements}
This study was supported by grant BFU2011-30063 (Spanish government).

\bibliography{document}
\bibliographystyle{unsrtnat}

\end{document}